\documentclass[preprint,aps,showpacs,preprintnumbers,amsmath,amssymb,11pt]{revtex4}
\usepackage{natbib}%
\usepackage{graphicx}
\usepackage{dcolumn}
\usepackage{bm}
\usepackage{psfrag}

\begin{document}
\newcommand{\be}{\begin{equation}}\newcommand{\ee}{\end{equation}}
\newcommand{\bea}{\begin{eqnarray}}\newcommand{\eea}{\end{eqnarray}}
\newcommand{\bc}{\begin{center}}\newcommand{\ec}{\end{center}}
\def\no{\nonumber}
\def\eq#1{Eq. (\ref{#1})}\def\eqeq#1#2{Eqs. (\ref{#1}) and  (\ref{#2})}
\def\lsim{\raise0.3ex\hbox{$\;<$\kern-0.75em\raise-1.1ex\hbox{$\sim\;$}}}
\def\gsim{\raise0.3ex\hbox{$\;>$\kern-0.75em\raise-1.1ex\hbox{$\sim\;$}}}
\def\slash#1{\ooalign{\hfil/\hfil\crcr$#1$}}
\def\eff{\mbox{\tiny{eff}}}
\def\order#1{{\mathcal{O}}(#1)}
\def\pppm{B^0\to\pi^+\pi^-}
\def\pzpz{B^0\to\pi^0\pi^0}
\def\pppz{B^0\to\pi^+\pi^0}
\preprint{ }
\title{Nonet symmetry in $\eta$, $\eta^{\prime}$  and 
$B\to K\eta,K\eta^{\prime}$ decays}
\author{
T. N. Pham}
\affiliation{
 Centre de Physique Th\'{e}orique, CNRS \\ 
Ecole Polytechnique, 91128 Palaiseau, Cedex, France }
\date{\today}
\begin{abstract}
The nonet symmetry scheme seems to describe rather well the masses and
$\eta-\eta^{\prime}$ mixing angle of the ground state pseudo-scalar 
mesons. It is expected that  nonet symmetry should also be valid 
for the matrix elements of the pseudo-scalar density operators 
which play an important role in charmless two-body $B$ decays with 
$\eta$ or  $\eta^{\prime}$ in the final state. Starting from the 
divergences of the $SU(3)$ octet and singlet 
axial vector currents, we show that  nonet symmetry  for the 
pseudo-scalar mass term implies  nonet symmetry for the
pseudo-scalar density operators. In this nonet symmetry scheme, we find
that the branching ratio $B\to PP,PV$, with $\eta$ in the final state 
agrees well with data, while those with $\eta'$  are underestimated, 
but by increasing the $B\to \eta'$ form factor
by $40-50\%$, one could explain the tree-dominated $B^{-}\to
\pi^{-}\eta'$ and $B^{-}\to \rho^{-}\eta'$  measured branching 
ratios. With this increased form factor and with only a  moderate 
annihilation contribution, we are able to obtain $62\times 10^{-6}$ 
for the penguin-dominated  $B^{-}\to K^{-}\eta'$ branching 
ratios, quite close to the measured value. This supports the predicted  
value for the $B\to \eta'$ form factor in PQCD and light-cone 
sum rules approach. A possible increase by $15\%$ of 
$\langle 0|\bar{s}\,i\gamma_5 s|s\bar{s}\rangle $ for $\eta_{0} $ would 
bring the predicted $B^{-}\to K^{-}\eta'$ branching ratio to
$69.375\times 10^{-6}$,  very close to experiment. 
\end{abstract}
\maketitle
\section{INTRODUCTION}
Unlike the low-lying vector mesons where the flavor diagonal $1^{-}$
$q\bar{q}$
states are eigenstate because of the OZI selection rule, the $0^{-}$
pseudo-scalar $q\bar{q}$ state can mix with each other. Since QCD
interactions through the exchange of gluons are flavor-independent, one
expects the wave function  for the pseudo-scalar nonet  also 
flavor-independent in the limit of vanishsing current quark mass 
($m_{q}\to 0, q=u,d,s$) and the 
$\eta$ and $\eta^{\prime}$ can be described as two linear combinations of the
$q\bar{q}$ state, the SU(3) singlet $\eta_{0}$ and the SU(3) octet 
$\eta_{8}$ which mix with each other through a small $SU(3)$ symmetry
breaking mixing parameter.
In fact, with  $m_{u}$ and $m_{d}$ $\ll m_{s}$ , $m_{s}\ll
\Lambda_{\rm QCD}$ , and because of the $U(1)$ QCD-anomaly, the $\eta_{0}$
mass is much larger compared to the $\eta_{8}$ mass, the 
$\eta-\eta^{\prime}$ mixing angle  is 
$O(m_{s}/\Lambda_{\rm QCD})$ so that
the physical $\eta$ and $\eta^{\prime}$ are almost  pure
$\eta_{8}$ and $\eta_{0}$ eigenstate respectively, in contrast with the ideal
mixing for the $1^{-}$ low-lying vector meson states. Another features 
of the $0^{-}$ $q\bar{q}$ nonet is that, because of the spontaneous 
breakdown of $SU(3)\times SU(3)$ symmetry, the octet mesons are 
 massless Goldstone bosons in the limit of vanishing current quark 
mass. This simplifies considerably the description of the ground state 
pseudo-scalar meson system. As shown in\cite{Donoghue}, a rather 
accurate description of the mass and mixing
angle in the $\eta-\eta^{\prime}$ system is obtained by adding an $U(1)$
QCD-anomaly  term for the $\eta_{0}$ mass in the nonet pseudo-scalar 
mass matrix. This 
mass matrix is generated by the quark mass term and is the leading
term in the large $N_{c}$ expansion while higher order terms 
in the chiral Lagrangian\cite{Gasser}
is $O(1/N_{c})$ and is thus suppressed in the large $N_{c}$ limit.
This justifies the nonet symmetry mass term for the pseudo-scalar 
mass matrix.
Vice-versa, from the nonet symmetry value  for the off-diagonal mass term
$<\eta_{0}|H_{\rm SB}|\eta_{8}>$, where 
$H_{\rm SB}= m_{s}\,\bar{s}s +m_{u}\,\bar{u}u + m_{d}\,\bar{d}d$ one  
would get a mixing angle $\theta= -18^{\circ}$ in good agreement with
a value $\theta \approx -(22 \pm 3)^{\circ}$ in \cite{Donoghue}, or 
$\theta \approx -(18.4 \pm 2)^{\circ}$ in \cite{Pham}  
and a similar value $\theta\approx -(17-20)^{\circ}$ \cite{Ball}
obtained from the two-photon decay width of $\eta$ and $\eta^{\prime}$. 
However, if we use the Gell-Mann-Okubo(GMO) mass formula
for the octet mass $m_{8}^{2}$,  we would have,
in terms of  the $\eta-\eta^{\prime}$ mixing angle $\theta$
\be
m_{\eta}^{2} = m_{8}^{2} - {\tan \theta}^{2}\,(m_{\eta'}^{2}- m_{8}^{2})
\label{etamass}
\ee
which, for $\theta = -18^{\circ}$ gives $m_{\eta}= 483\,\rm MeV $, 
about $60\,\rm MeV $ below experiment. Thus  the $\eta-\eta^{\prime}$
mixing which contributes to $L_{7}$  in \cite{Gasser} has driven the 
$m_{\eta} $ below the GMO value by $63\,\rm MeV $. This is also 
the case with a nonet mass matrix in the quark basis\cite{Georgi,Gerard} 
which has a large $\eta-\eta^{\prime}$ mixing and an upper bound 
for the $\eta$ mass far below experiment. The  higher order 
terms $L_{4},L_{5}, L_{6}, L_{8}$ and chiral logarithms obtained in 
Ref. \cite{Gasser} shift $m_{\eta} $ 
upward by a similar amount  with the result that the $\eta$ mass 
is very close to the GMO value, in agreement with experiment. Similar 
result is also obtained in \cite{Gerard} more recently. 
Thus, nonet symmetry seems to be a good approximation for 
the $0^{-}$ nonet mass term. One could then go further and try to see if 
the matrix elements of the pseudo-scalar density local operator 
{\em e.g.} $\bar{s}\,i\gamma_{5}\,s$ could also satisfies nonet symmetry.
This  will allow a simple
calculation of the penguin matrix elements in the charmless two-body 
decays of $B$ meson with  $\eta$ or  $\eta^{\prime}$ in the final states. 
In this paper we will  use the divergence equation for the octet
and singlet axial vector current to show that  nonet symmetry scheme
for the mass term implies  nonet symmetry for the pseudo-scalar density
$\bar{q}\,i\gamma_{5}\,q$ for  $\eta$ and $\eta^{\prime}$. The basic idea 
is to include in the matrix elements 
of the axial vector current and its divergence the $\eta_{0,8}$ pole 
contribution which  will add  the mixing mass term
 $<\eta_{0}|H_{\rm SB}|\eta_{8}>$ 
to  the divergence equation and allows us to obtain the nonet
symmetry expression for the matrix element of the pseudo-scalar density
operators between the vacuum and $\eta_{0,8}$ . In the next 
section we will first derive a 
divergence equation for the $\bar{u}\,\gamma_{\mu}\gamma_{5}\,u$ 
and $\bar{s}\,\gamma_{\mu}\gamma_{5}\,s$  axial vector current, in the 
presence of the $SU(3)\times SU(3)$-breaking $H_{\rm SB} $ current quark
mass term. Section III is an analysis of 
$B^{-}\to P\eta$ and $B^{-}\to P\eta^{\prime}$ ,$P= K^{-},\pi^{-}$ in 
 QCD Factorization (QCDF) with  nonet symmetry for the 
pseudo-scalar density and $B\to \eta$ and  $B\to \eta^{\prime}$ transition
form factors. We find that the  branching ratio for modes  
with $\eta$ in the final state 
agrees well with data, while those with $\eta'$ in the final state are 
underestimated. We then  increase the $B\to \eta'$ form factor
by $40-50\%$, to bring the tree-dominated $B^{-}\to \pi^{-}\eta'$ and  
$B^{-}\to \rho^{-}\eta'$ to the measured values. The increased form 
factor is then used to obtain a branching ratio close to data
for the penguin-dominated $B^{-}\to K^{-}\eta'$ decay.
\section{Pseudo-scalar density matrix element and nonet symmetry }
Let $|\eta_{0}>, |\eta_{8}>$ be the SU(3) singlet and octet
eigenstate  of the $I=0$, pseudo-scalar nonet in the absence of the
$SU(3)$ symmetry breaking quark mass term $H_{\rm SB}$, in terms of the
flavor diagonal $q \bar{q}$ component:
\bea
&& |\eta_{0} > =  (| u \bar{u}+ d \bar{d} + s \bar{s}>)/\sqrt{3} , \nonumber\\
&&  |\eta_{8} > = (| u \bar{u}+ d \bar{d} - 2\,s \bar{s}>)/\sqrt{6} . 
\label{eta0}
\eea
Consider now  the matrix element of the axial vector current matrix element
$\bar{u}\,\gamma_{\mu}\gamma_{5}u$ and  $\bar{s}\,\gamma_{\mu}\gamma_{5}s$
between the vacuum and $\eta_{0} $ and  $\eta_{8} $ :
\bea
&& <0|\bar{u}\,\gamma_{\mu}\gamma_{5}u|\eta_{0} > = i\,f_{u}\,p_{\mu}/\sqrt{3},  \nonumber\\
&&  <0|\bar{u}\,\gamma_{\mu}\gamma_{5}u|\eta_{8} > = i\,f_{u}\,p_{\mu}/\sqrt{6}.
\label{Au}
\eea
and 
\bea
&& <0|\bar{s}\,\gamma_{\mu}\gamma_{5}s|\eta_{0} > = i\,f_{s}\,p_{\mu}/\sqrt{3} , \nonumber\\
&&  <0|\bar{s}\,\gamma_{\mu}\gamma_{5}s|\eta_{8} > = -2\,i\,f_{s}\,p_{\mu}/\sqrt{6}.
\label{As}
\eea
where $f_{u}$ and $f_{s}$ are defined as the decay constant of $u\bar{u}$
and $s\bar{s}$ state respectively. Except for the momentum dependence 
factor $p_{\mu}$, the above axial vector current matrix elements 
depend on the same $f_{u}$ and $f_{s}$ according to  nonet symmetry scheme
with identical $q\bar{q}$ spatial wave function in 
$\eta_{0}$ and $\eta_{8}$\cite{Donoghue}, but 
$f_{s}$ could be different from $f_{u}$
by an $SU(3)$ breaking $s$-quark mass term. 
The  octet $A_{8\,\mu}$ and singlet $A_{0\,\mu}$ axial vector current 
matrix elements
between the vacuum and $\eta_{8},\eta_{0}$  are then given by:
\bea
&& <0|A_{\mu 8}|\eta_{8}> = \frac{(f_{u}+ f_{d} + 4\,f_{s})}{6}\,p_{\mu},
\nonumber\\
&& <0|A_{\mu 0}|\eta_{0}> = \frac{(f_{u}+ f_{d} + f_{s})}{3}\,p_{\mu}.
\label{A08}
\eea
($p_{\mu}$ is the 4-momentum of $\eta_{0}$ and $\eta_{8}$ .
Similar, for other members  of the $SU(3)$ octet, we have 
$f_{\pi}$ and $f_{K}$ for $\pi^{+}=u \bar{d}$,  $K^{+}=u \bar{s}$
meson respectively. Assuming each $s$-quark contributes to the decay
constant a symmetry breaking term $\epsilon$, to first order in
$\epsilon$, (Rewriting $f_{q\bar{q}}= f_{q}$),  we have\cite{Pham1}:
\bea
&& f_{\pi} = f_{u\bar{d}}\approx f_{u} , \quad     \nonumber \\
&& f_{K}= f_{u\bar{s}}= (1 + \epsilon)\,f_{u\bar{d}} \quad , \nonumber \\
&& f_{s}= (1 + 2\,\epsilon)\,f_{u}\approx (1 + \epsilon)\,f_{K}.
\eea
The usual way to obtain the pseudo-scalar density matrix elements is to 
take the divergence of the axial 
vector current between the vacuum and the pseudo-scalar meson
octet. For example, taking the matrix elements of
$\bar{u}\,i\gamma_{5}\,d$, $\bar{u}\,i\gamma_{5}\,s$, 
$(\bar{u}\,i\gamma_{5}\,u - (\bar{d}\,i\gamma_{5}\,d)$ 
between the vacuum and $\pi^{+}$, $K^{+}$, $\pi^{0}$, respectively,
we have:
\bea
&& f_{\pi}B_{0}(m_{u} + m_{d}) = (m_{u}+ m_{d})\langle 0|\bar{u}\,i \gamma_5
 d|u\bar{d}\rangle ,\nonumber\\
&& f_{K}B_{0}(m_{u} + m_{s}) = (m_{u}+ m_{s})\langle 0|\bar{u}\,i \gamma_5
 s|u\bar{s}\rangle .
\label{pi+}
\eea
and for $\pi^{0}$ :
\be
 f_{u}B_{0}(m_{u} + m_{d}) = (m_{u}+ m_{d})\langle 0|\bar{u}\,i \gamma_5
 u|u\bar{u}\rangle . 
\label{pi0}
\ee
where the $\pi$ and $K$ meson masses are the usual expressions in terms
of $B_{0}$ and the current quark mass\cite{Gasser,Donoghue}. The expression
for $\pi^{0}$ is obtained by putting:\cite{Isola}
\be
\langle 0|\bar{u}\,i \gamma_5
 u|\pi^{0}\rangle = -\langle 0|\bar{d}\,i \gamma_5 d|\pi^{0}\rangle .
\ee
Apart from the difference in $f_{\pi}$ and $f_{K}$, we see that 
the above pseudo-scalar density matrix element in Eq.(\ref{pi+}) and 
Eq.(\ref{pi0}) satisfies SU(3) symmetry. We will see below that to have
nonet symmetry for the pseudo-scalar density matrix element between
the vacuum and $\eta_{0,8}$, the pole term in the divergence equation
must be included.
We now consider the divergence of the $I=0$  
 $A_{{\rm n}\,\mu}$  and $A_{{\rm s}\,\mu}$ axial vector current:
\bea
&& A_{{\rm n}\,\mu}= (\bar{u}\,\gamma_{\mu}\gamma_{5}u +
\bar{d}\,\gamma_{\mu}\gamma_{5}d), \quad  \nonumber \\
&& A_{{\rm s}\,\mu}= \bar{s}\,\gamma_{\mu}\gamma_{5}s. \quad 
\label{Ans}
\eea
The divergence is given by:
\bea
&&\partial A_{\rm n} = 2(m_{u}\bar{u} i \gamma_5 u + 
m_{d}\bar{d} i \gamma_5 d)  + 2\frac{\alpha_s}{4\pi} 
G\,\tilde{G} ,\quad \nonumber \\
&&\partial A_{\rm s}  = 2 m_s\bar{s} i \gamma_5 s + 
\frac{\alpha_s}{4\pi} G\,\tilde{G} . \quad 
\label{dAns}
\eea
 The matrix elements of $\partial A_{\rm n} $ and $\partial A_{\rm s}$
between the vacuum and $\eta_{0,8}$ are given by:
\bea
\kern -0.3cm \langle 0|\partial A_{\rm n}|\eta_{0}\rangle\kern -0.2cm &=&\kern -0.2cm 2m_{u}\langle 0|\bar{u}\, i
 \gamma_5 u|\eta_{0}\rangle + 2m_{d}\langle 0|\bar{d}\, i \gamma_5
 d|\eta_{0}\rangle , \label{dA0u} \\
&+&2\langle 0|\frac{\alpha_s}{4\pi}G\,\tilde{G}|\eta_{0}\rangle .\quad \nonumber \\
\langle 0|\partial A_{\rm s}|\eta_{0}\rangle\kern -0.2cm &=&\kern -0.2cm 2m_{s}\langle 0|\bar{s}\, i
 \gamma_5 s|\eta_{0}\rangle    + \langle 0|\frac{\alpha_s}{4\pi}G\,\tilde{G}|\eta_{0}\rangle . \quad 
\label{dA0s}
\eea
and for $\eta_{8}$ :
\bea
\kern -0.3cm \langle 0|\partial A_{\rm n}|\eta_{8}\rangle\kern -0.2cm &=&\kern -0.2cm 2m_{u}\langle 0|\bar{u}\, i \gamma_5 u|\eta_{8}\rangle + 2m_{d}\langle 0|\bar{d}\, i \gamma_5
 d|\eta_{8}\rangle , \nonumber\\
& +& 2\langle 0|\frac{\alpha_s}{4\pi}G\,\tilde{G}|\eta_{8}\rangle \quad
,\label{dA8u}\\
\langle 0|\partial A_{\rm s}|\eta_{8}\rangle\kern -0.2cm &=& \kern -0.2cm 2m_{s}\langle 0|\bar{s}\, i \gamma_5 s|\eta_{8}\rangle    + \langle 0|\frac{\alpha_s}{4\pi}G\,\tilde{G}|\eta_{8}\rangle.  \quad 
\label{dA8s}
\eea
In the limit $m_{u}=m_{d}=0$ , since the l.h.s of Eq.(\ref{dA8u}) is 
$f_{u}\,m_{8}^{2}$, the matrix element
$2\langle 0|\frac{\alpha_s}{4\pi}G\,\tilde{G}|\eta_{8}\rangle $ on the r.h.s
is   $O(m_{8}^{2})$ and is given by the $\eta_{0}$ pole contribution. 
We now evaluate  Eq.(\ref{dA0u}-\ref{dA0s}) and Eq.(\ref{dA8u}-\ref{dA8s}) 
with the pole terms included using the nonet symmetry expressions 
for  $m_{0,8}^{2}$ and $m_{08}^{2}$\cite{Donoghue} :
\bea
m_{8}^{2} &=& B_{0}\frac{2}{3}\,(2m_{s} + \hat{m}), \nonumber\\
m_{0}^{2} &=& \bar{m}_{0}^{2} + B_{0}\frac{2}{3}(m_{s} + 2\hat{m}), \nonumber\\
m_{08}^{2} &=& B_{0}\frac{2}{3}\sqrt{2}(-m_{s} + \hat{m}) . 
\label{mass}
\eea
in standard notation\cite{Gasser} ($\hat{m}=(m_{u}+ m_{d})/2$).
At the $\eta_{0}$ and $\eta_{8}$ mass, $p^{2}=m_{0}^{2}$ and $p^{2}=m_{8}^{2}$
in the l.h.s of Eq.(\ref{dA0u}-\ref{dA0s}) and Eq.(\ref{dA8u}-\ref{dA8s}) 
respectively. As mentioned above, since $m_{u,d}\ll m_{s}$, $SU(3)$ is broken
and the $\eta_{0,8}$ pole will contribute to both the l.h.s and r.h.s
of Eq.(\ref{dA0u}-\ref{dA0s}) and Eq.(\ref{dA8u}-\ref{dA8s}) . The pole 
terms on the r.h.s come from the  QCD-anomaly matrix element 
$\langle 0|\frac{\alpha_s}{4\pi}G\,\tilde{G}|\eta_{0}\rangle $
and 
$\langle 0|\frac{\alpha_s}{4\pi}G\,\tilde{G}|\eta_{8}\rangle $ induced
by $SU(3)$-breaking $\eta_{0}-\eta_{8}$ mixing mass term $m_{08}^{2}$ .
The presence of the $\eta_{0,8}$ pole term is important, since its 
contribution is the same order as the current-quark mass terms in 
$m_{0,8}^{2}$ . Indeed had  we dropped the 
$\eta_{0,8}$ pole term we would run into contradiction with the divergence
equation. To obtain the pseudo-scalar density matrix elements, 
let us bring the $p^{2}$-dependence pole term 
in the l.h.s to  the r.h.s of Eq.(\ref{dA0u}-\ref{dA0s}) and 
Eq.(\ref{dA8u})-(\ref{dA8s}). Putting  $f_{u}= f_{d}$ and 
$\langle 0|\bar{u}\, i\gamma_5 u|\eta_{0,8}\rangle
= \langle 0|\bar{d}\, i\gamma_5 d|\eta_{0,8}\rangle $ ,  we 
find, for $\eta_{0}$ :
\bea
\kern -0.3cm f_{u}\frac{1}{\sqrt{3}}(\bar{m}_{0}^{2} + B_{0}\frac{2}{3}(m_{s} + 
2{\hat  m}))&=& f_{u}\frac{1}{\sqrt{3}}\bar{m}_{0}^{2} - 
 \kern -0.1cm f_{u}\frac{1}{\sqrt{6}}(B_{0}\frac{2\sqrt{2}}{3}({\hat m}
-m_{s}))+2\frac{1}{\sqrt{3}}{\hat m}\langle 0|\bar{u}\, i\gamma_5
 u|u\bar{u}\rangle ,\label{uu0} \\
\kern -0.1cm f_{s}\frac{1}{\sqrt{3}}(\bar{m}_{0}^{2} + B_{0}\frac{2}{3}(m_{s} + 2{\hat m}))
&=& f_{s}\frac{1}{\sqrt{3}}\bar{m}_{0}^{2} - 
 \kern -0.1cm f_{s}\frac{2}{\sqrt{6}}B_{0}\frac{2\sqrt{2}}{3}({\hat m}
-m_{s}) +2\frac{1}{\sqrt{3}}m_{s}\langle 0|\bar{s}\, i\gamma_5 s|s\bar{s}\rangle. 
\label{ss0}
\eea
and. similarly, for $\eta_{8}$ :
\bea
\kern -0.3cm f_{u}\frac{1}{\sqrt{6}}( B_{0}\frac{2}{3}(2m_{s} + 
{\hat  m}))&=& -f_{u}\frac{1}{\sqrt{3}}B_{0}\frac{2\sqrt{2}}{3}({\hat m}
-m_{s})
+2\frac{1}{\sqrt{6}}{\hat m}\langle 0|\bar{u}\, i\gamma_5
 u|u\bar{u}\rangle ,\label{uu8} \\
\kern -0.3cm -f_{s}\frac{2}{\sqrt{6}}( B_{0}\frac{2}{3}(2m_{s} + {\hat m}))
&=&  -f_{s}\frac{1}{\sqrt{3}}B_{0}\frac{2\sqrt{2}}{3}({\hat m}
-m_{s})
 -2\frac{2}{\sqrt{6}}m_{s}\langle 0|\bar{s}\, i\gamma_5 s|s\bar{s}\rangle .
\label{ss8}
\eea

Comparing the l.h.s and the r.h.s of Eq.(\ref{uu0}) and Eq.(\ref{ss0}), we
get the pseudo-scalar density matrix element for $\eta_{0}$:
\bea
&&\langle 0|\bar{u}\,i\gamma_5 u|u\bar{u}\rangle = B_{0}f_{u},\\
\label{u0}
&&\langle 0|\bar{s}\,i\gamma_5 s|s\bar{s}\rangle = B_{0}f_{s}.
\label{s0}
\eea
Similarly, by comparing l.h.s and the r.h.s of Eq.(\ref{uu8}) 
and Eq.(\ref{ss8}), we get the same expression for the pseudo-scalar 
density matrix element, but in $\eta_{8}$ .

We have shown  that, by including the $\eta_{0}$ and $\eta_{8}$ pole in the
divergence equations, and by using the nonet symmetry expressions for
the current quark mass contributions to the $\eta_{0}$ and $\eta_{8}$
mass, the pseudo-scalar density operators matrix elements
between $\eta_{0}$ and $\eta_{8}$ can be obtained by  nonet symmetry
and  quark counting rule. Like 
 the matrix elements  $\langle 0|\bar{u}\,i\gamma_5 d|\pi^{+}\rangle  $,
$\langle 0|\bar{u}\,i\gamma_5 u|\pi^{0}\rangle $ and 
$\langle 0|\bar{u}\,i\gamma_5 s|K^{+}\rangle $, they are given by 
the parameter $B_{0}$ and the decay constant involved. 
Experimentally, from the known value of the $\eta-\eta^{\prime}$
mixing angle, $\theta =(-20\pm 2)^{\circ}$ , one has
 $m_{08}^{2}= -(0.81\pm 0.05)\,m_{K}^{2}$ to be compared
with the nonet symmetry value of 
$m_{08}^{2}\simeq -0.90\,m_{K}^{2}$\cite{Donoghue}, 
we expect  nonet symmetry for the pseudo-scalar
density matrix elements in $\eta -\eta^{\prime}$ valid to this accuracy. 
Since the octet $m_{8}^{2}$ mass gets about $15\%$ increase from
higher order terms $L_{4},L_{5}, L_{6}, L_{8}$ and chiral 
logarithms\cite{Gasser},  Eqs.(\ref{uu8}-\ref{ss8}) show that 
$\langle 0|\bar{s}\,i\gamma_5 s|s\bar{s}\rangle $ in $\eta$ will be 
increased by
a similar amount. Note that the r.h.s of Eqs.(\ref{uu8}-\ref{ss8}) 
gets this increase from higher order terms in the pole and other 
terms.  Higher order $SU(3)$ breaking 
contribution to the singlet $m_{0}^{2}$ mass is
not known, but if we assume a similar $15\%$ increase from the nonet 
value in Eq.(\ref{mass}), $\langle 0|\bar{s}\,i\gamma_5
s|s\bar{s}\rangle $ in $\eta_{0}$ will also be increased by a similar amount.
This could be another source of enhancement for the  $B\to K\eta'$ 
branching ratio, as found below.
We note that it might be possible  to obtain the 
pseudo-scalar density matrix elements  in Eqs.(\ref{uu0}-\ref{ss0})
and  Eqs.(\ref{uu8}-\ref{ss8}) using the known  values of $m_{0,8}^{2}$
and \break $m_{08}^{2}$ , but because
of the precise dependence on quark mass is not known and 
experimental errors involved, the physical interpretation of the 
result will be lost. We would like to stress that in our derivation, 
the anomaly contribution to the $\eta_{0}$ mass has been included
in the divergence equation, thus the enhancement factor for 
$\langle 0|\bar{s}\,i\gamma_5 s|\eta_{0}\rangle $
suggested in \cite{Kou} would  have the origin  elsewhere. With the 
pseudo-scalar density matrix elements given above and  nonet symmetry for
the $B\to \eta,\eta^{\prime}$ transition form factors, we shall now 
compute the $B^{-}\to K^{-}\eta, K^{-}\eta^{\prime}$ and 
$ B^{-}\to \pi^{-}\eta,\pi^-\eta^{\prime}$
decay branching ratios in QCD Factorization(QCDF). 
\section{$B^{-}\to K^{-}(\eta,\eta^{\prime})$ AND 
 $ B^{-}\to \pi^{-}(\eta,\eta^{\prime})$ DECAY IN QCD FACTORIZATION}
The $B\to M_{1} M_{2}$ decay amplitude in QCDF is given by\cite{QCDF1,QCDF2}: 
\be
 {\cal A}(B \rightarrow M_1 M_2)=
 \frac{G_F}{\sqrt{2}}\sum_{p=u,c}V_{pb}V^{*}_{ps}\times 
 \left( -\sum_{i=1}^{10} a_i^p
   \langle M_1 M_2 \vert O_i \vert B \rangle_H + 
 \sum_{i}^{10} f_B f_{M_1}f_{M_2} b_i \right ),
\label{BMM}
\ee
where the QCD coefficients  $a_{i}^{p}$ contain the vertex corrections,
penguin corrections, and hard spectator scattering contributions, 
the hadronic matrix elements $ \langle M_1 M_2 \vert O_i \vert B
\rangle_H $  of the tree and penguin operators $O_{i}$ are given 
by factorization model\cite{Ali,Zhu1}, $b_{i}$ are annihilation
 contributions. The values for $a_{i}^{p}$,$p=u,c$ , computed from 
the expressions in \cite{QCDF1,QCDF2} at the renormalization 
scale $\mu=m_{b}$, with $m_{b}=4.2\,\rm GeV$ are:
\bea
&& a_{4}^{c}=-0.033 - 0.013\,i + 0.0009\,\rho_{H},\nonumber \\
&& a_{4}^{u}=-0.027 - 0.017\,i + 0.0009\,\rho_{H},\nonumber \\
&& a_{6}^{c}=-0.045 - 0.003\,i ,\quad a_{6}^{u}=-0.042 - 0.013\,i ,\nonumber \\
&& a_{8}^{c}=-0.0004 - 0.0001\,i ,\quad  a_{8}^{u}= 0.0004 - 0.0001\,i ,\nonumber \\
&& a_{10}^{c}=-0.0011 - 0.0001\,i - 0.0006\,\rho_{H} ,\nonumber \\
&& a_{10}^{u}=-0.0011 + 0.0006\,i - 0.0006\,\rho_{H}.
\label{aiuc}
\eea
for $i=4,6,8,10$. For other coefficients, $a_{i}^{u}=a_{i}^{p}=a_{i}$ :
\bea
&& a_{1}= 1.02 + 0.015\,i -0.012\,\rho_{H},\nonumber \\
&& a_{2}= 0.156 - 0.089\,i + 0.074\,\rho_{H}, \nonumber \\
&& a_{3}= 0.0025 + 0.0030\,i - 0.0024\,\rho_{H}, \nonumber \\
&& a_{5}=-0.0016 - 0.0034\,i + 0.0029\,\rho_{H}, \nonumber \\
&& a_{7} =-0.00003 - 0.00004\,i - 0.00003\,\rho_{H},\nonumber \\
&& a_{9} = -0.009 - 0.0001\,i + 0.0001\,\rho_{H}. \nonumber \\
\label{ai}
\eea
where the complex parameter $\rho_{H}\exp(i\phi_{H})$ represents the 
end-point singularity contribution in the hard-scattering
corrections $X_{H}=(1 +\rho_{H}\exp(i\phi_{H}))\,\ln(\frac{m_{B}}{\Lambda_{h}})$\cite{QCDF1,QCDF2}
(we have put the phase $\phi_{H}=0$ in the above expressions).

For the annihilation terms, we have:
\bea
&& b_{2}= -0.0038 - 0.0065\,\rho_{A} - 0.0018\rho_{A}^{2} ,\nonumber \\
&& b_{3}=  -0.0065 - 0.0150\, \rho_{A} - 0.0085\, \rho_{A}^{2}, \nonumber \\
&& b_{3}^{ew}= -0.00011 - 0.00015\,\rho_{A} + 0.000003\,\rho_{A}^{2}. 
\label{b3}
\eea
where $b_{i}$ are evaluated with the factor $f_{B}f_{M_{1}}f_{M_{2}}$
included and $\rho_{A}$ , like $\rho_{H}$, appears in the divergent 
annihilation term 
$X_{A}=(1 +\rho_{A}\exp(i\phi_{A}))\,\ln(\frac{m_{B}}{\Lambda_{h}})$.

For the CKM matrix elements, since the inclusive and exclusive data on
$|V_{ub}|$ differ by a large amount and  the higher inclusive  data
exceeds the unitarity limit for 
$R_{b} = |V_{ud}V_{ub}^{*}|/|V_{cd}V_{cb}^{*}| $ with the current value
$\sin(2\beta)=0.687\pm 0.032 $\cite{PDG}, we shall determine $|V_{ub}|$ from
the more precise $|V_{cb}|$ data. We have\cite{CKM}:
\be
\kern -0.5cm \vert V_{ub}\vert=  \frac{\vert V_{cb}V_{cd}^{*}\vert}{\vert V_{ud}^{*}\vert} \vert  \sin \beta 
\sqrt{1+\frac{\cos^2 \alpha}{\sin^2 \alpha}} .
\label{Vub}
\ee
 With $\alpha=(99^{+13}_{-9})^{\circ}$\cite{PDG} and 
$\vert V_{cb}\vert = (41.78\pm 0.30\pm 0.08)\times 10^{-3}$ \cite{Barberio}, we find
\be
\vert V_{ub}\vert = 3.60\times 10^{-3}.
\label{Vub1}
\ee
in good agreement with the exclusive data in the range
$\vert V_{ub}\vert = 3.33 -3.51$\cite{Barberio} . 
The recent measurements of the $B_{s}-\bar{B}_{s}$ mixing 
also allow the extraction of $|V_{td}|$ from $B_{d}-\bar{B}_{d}$
mixing data. The  current determination\cite{Abulencia} gives
$|V_{td}/V_{ts}|= (0.208^{+0.008}_{-0.006})$ which in turn can be used to
determined the  angle $\gamma$ from the unitarity relation\cite{CKM}:
\be
\kern -0.5cm \vert V_{td}\vert= \frac{\vert V_{cb}V_{cd}^{*}\vert}{\vert V_{tb}^{*}\vert} \vert  \sin \gamma
\sqrt{1+\frac{\cos^2 \alpha}{\sin^2 \alpha}}.
\label{Vtd}
\ee
with $|V_{tb}|=1 $, we find $\gamma = 66^{\circ} $
which implies an angle $\alpha = 91.8^{\circ}$, in good agreement 
with the value found in the current UT-fit value of $(88\pm 16)^{\circ}$
\cite{Brown}. In the following in our $B$ decay calculations, we shall
use the unitarity triangle values for $\vert V_{ub}\vert $ and $\gamma$. For 
 other hadronic parameters we use the values in 
Table 1  of \cite{QCDF2} and take $m_{s}(\rm 2\,GeV)=80\,\rm MeV$,
$ f_{u}=f_{\pi}$, $f_{s}= f_{\pi}\left(1 + 2(\frac{f_{K}}{f_{\pi}}-1)\right)$.
For the $B\to \pi$ and $B\to K$ transition form factor, we use the current
light-cone sum rules central value\cite{Zwicky}~:
\be
F^{B\pi}_{0}(0)= 0.258,\quad   F^{BK}_{0}(0) = 0.33
\label{FBK}
\ee
With $\eta-\eta^{\prime}$ mixing angle $\theta = -20^{\circ}$, we have 
\bea
&&|\eta \rangle = (0.58(|u\bar{u}\rangle + |d\bar{d}\rangle) -
0.57|s\bar{s}\rangle), \quad \nonumber \\
&&|\eta^{\prime} \rangle = (0.40(|u\bar{u}\rangle + |d\bar{d}\rangle) + 0.82|s\bar{s}\rangle).
\label{content}
\eea
From Eq.(\ref{content}, we find~:
\be
F^{B\eta}_{0}(0) = 0.58\,F^{B\pi}_{0}(0), \  F^{B\eta'}_{0}(0) = 0.40\,F^{B\pi}_{0}(0).
\label{FBeta}
\ee
The $B\to K(\eta',\eta)$ decay amplitude  can now be obtained
from  the factorization formula for the hadronic matrix elements 
in Eq.(\ref{BMM}) with  the pseudo-scalar density matrix
element obtained in Eq.(\ref{u0}) and the form factors given above. We
have
\bea
&&\langle 0|\bar{s}\,i\gamma_5 s|\eta\rangle=C_{\eta}\,B_{0}f_{s}, \nonumber\\
&&\langle 0|\bar{s}\,i\gamma_5 s|\eta'\rangle=C_{\eta'}\,B_{0}f_{s}.
\label{O6}
\eea
where $B_{0}= m_{K}^{2}/(m_{s}+ {\hat m})$ and $C_{\eta}=-0.57$, 
$C_{\eta'}= 0.82 $, the fraction of $s\bar{s}$ state in $\eta$ and $\eta'$
respectively. This contributes to the $O_{6}$ matrix element a term
$f_{s}r_{\chi}^{K}$, with $r_{\chi}^{K}= 2m_{K}^{2}/(m_{b} + {\hat
  m})\,(m_{s}+ {\hat m})$, similar to that
in $\bar{B}^{0}\to K^{-}\pi^{+}$ decay, except that in $B^{-}\to K^{-}\eta$
and $B^{-}\to K^{-}\eta'$, the $O_{6}$ matrix element is enhanced by a factor
$f_{s}/f_{K}$. In this way, the decay amplitude in unit
of $\rm GeV$ are :
\bea
A(B^{-}\to \pi^{-}\pi^{0})&=& (0.110 + 0.204\,i)\times 10^{-7}, \nonumber\\    
A(\bar{B}^{0}\to K^{-}\pi^{+})&=&
 -(0.368 + 0.004\,i)\times 10^{-7}\,(F^{B\to \pi}_{0}(0)/0.258) \nonumber\\   
&&-(0.090 + 0.002\,i)\times 10^{-7}.
\label{pipi}
\eea
from which the branching ratios are, with $\rho_{H}=0$,$\rho_{A}=0.6$
(only the central values for the relevant parameters are used in the
calculations)
\bea
&&{\cal B}(B^{-}\to \pi^{-}\pi^{0})= 5.050\times 10^{-6} ,\nonumber\\ 
&&{\cal B}(\bar{B}^{0}\to K^{-}\pi^{+})= 18.249\times 10^{-6}.          
\label{brpipi}
\eea
in good agreement with the current measured branching ratios\cite{HFAG}
\bea
&&{\cal B}(B^{-}\to \pi^{-}\pi^{0})= 
 (5.7\pm 0.4\times 10^{-6} ,\nonumber\\ 
&&{\cal B}(\bar{B}^{0}\to K^{-}\pi^{+})=(19.04 \pm 0.6)\times 10^{-6}.         
\label{brexp}
\eea
 We note a sizable  annihilation contribution, given by the last term 
in Eq.(\ref{pipi}), is needed to produce a large 
${\cal B}(\bar{B}^{0}\to K^{-}\pi^{+})$ .
This is not surprising since annihilation contribution is also needed
to explain the large branching ratios of $B^+ \to \pi^+ K^{\ast 0}$
and $B^0 \to K^- \rho^+$ decay\cite{Zhu3}.
Our result also shows that the values $0.258$ for $F^{B\pi}_{0}(0) $ 
and $0.33$ for $F^{BK}_{0}(0)$ given above are reasonable. We will use 
these values in the calculation of the decay modes with $\eta$ 
and $\eta'$. We find
\bea
 A(B^{-}\to K^{-}\eta)&=& 
  -(0.283+0.032\,i)\times 10^{-7}\,(F^{B\to \eta}_{0}(0)/0.150)  \nonumber \\
&& +(0.317 + 0.080\,i)\times 10^{-7}\,(F^{B\to K}_{0}(0)/0.33) \nonumber \\
&& +(0.015 + 0.0004\,i)\times 10^{-7}.        
\label{ameta}
\eea
\bea
A(B^{-}\to K^{-}\eta') &=&
 -(0.192 +0.022\,i)\times 10^{-7}\,(F^{B\to \eta'}_{0}(0)/0.104) \nonumber\\
&& -(0.425 + 0.039\,i)\times 10^{-7}\,(F^{B\to K}_{0}(0)/0.33) \nonumber \\
&& -(0.111 + 0.003\,i)\times 10^{-7}.     
\label{ameta'}
\eea
where the last term in Eq.(\ref{ameta}) and Eq.(\ref{ameta'}) are the 
annihilation contributions($\rho_{H}=0, \rho_{A}=0.6$). The predicted 
branching ratios are then~:
\bea
&& {\cal B}(B^{-}\to K^{-}\eta)= 0.431\times 10^{-6} , \nonumber\\  
&& {\cal B}(B^{-}\to K^{-}\eta')= 48.263\times 10^{-6}.
\label{etabr}
\eea 
to be compared with the current experimental values\cite{HFAG}:
\bea
&& {\cal B}(B^{-}\to K^{-}\eta)= (2.2\pm 0.3)\times 10^{-6}, \nonumber\\  
&& {\cal B}(B^{-}\to K^{-}\eta')= (69.7^{+2.8}_{-2.7} )\times 10^{-6}.
\label{etaexp}
\eea
We see that the  ${\cal B}(B^{-}\to K^{-}\eta') $ is 
underestimated by about $30\%$, while the ${\cal B}(B^{-}\to K^{-}\eta)$ 
is very much suppressed, but because of large cancellation in 
the $ B^{-}\to K^{-}\eta $ amplitude due to the negative $s\bar{s}$ 
amplitude in the $\eta$ meson wave function, a precise prediction for 
${\cal B}(B^{-}\to K^{-}\eta) $ is more difficult. For  
$B^{-}\to K^{-}\eta'$, since $b_{3}$ contributes both to 
$\bar{B}^{0}\to K^{-}\pi^{+}$ and $B^{-}\to K^{-}\eta'$ decays, it is difficult
to adjust the annihilation term for $B^{-}\to K^{-}\eta'$ without 
overestimating the $\bar{B}^{0}\to K^{-}\pi^{+}$ branching
ratio. Another possibility is to increase the form factor 
$F^{B\to \eta'}_{0}(0) $ from the nonet symmetry value to bring the
predicted value closer to data. That this is the case can be seen by looking
at the $B^{-}\to \pi^{-}\eta'$ decays. We have:
\bea
 A(B^{-}\to \pi^{-}\eta)&= & (0.119+0.147\,i)\times 10^{-7}\,(F^{B\to \eta}_{0}(0)/0.150)  \nonumber \\
&& -(0.002 - 0.003\,i)\times 10^{-7}\,(F^{B\to \pi}_{0}(0)/0.258)\nonumber \\
 && -(0.004 - 0.003\,i)\times 10^{-7} .        
\label{pieta}
\eea
\bea
 A(B^{-}\to \pi^{-}\eta')& =&
 (0.081 +0.100\,i)\times 10^{-7}\,(F^{B\to \eta'}_{0}(0)/0.104)  \nonumber\\
&& +(0.008 - 0.002\,i)\times 10^{-7}\,(F^{B\to \pi}_{0}(0)/0.258) \nonumber \\
&& +(0.033 - 0.021\,i)\times 10^{-7} .     
\label{pieta'}
\eea
(the last term in the above amplitudes is the annihilation contributions).
This gives:
\bea
&& {\cal B}(B^{-}\to \pi^{-}\eta)= 3.388\times 10^{-6} , \nonumber\\  
&& {\cal B}(B^{-}\to \pi^{-}\eta')= 1.910\times 10^{-6} .
\label{etapibr}
\eea
comparing with the current measured branching ratios\cite{HFAG}:
\bea
&& {\cal B}(B^{-}\to \pi^{-}\eta)= (4.4\pm 0.4)\times 10^{-6} , \nonumber\\  
&& {\cal B}(B^{-}\to \pi^{-}\eta')= (2.6^{+0.6}_{-0.5})\times 10^{-6} .
\label{etapiexp}
\eea
we see that the predicted ${\cal B}(B^{-}\to \pi^{-}\eta) $ agrees more
or less with experiment, considering theoretical uncertainties in the
CKM parameters and in the $B\to \pi$ and $B\to K$ form factors.
 while ${\cal B}(B^{-}\to \pi^{-}\eta') $ is below the Babar
value of $ (4.0 \pm 0.8\pm 0.4)\times 10^{-6} $
\cite{HFAG}. Existing QCDF 
calculations\cite{Zhu4} also
underestimate ${\cal B}(B^{-}\to \rho^{-}\eta')$ by  a factor of
$\approx 2$ as seen from the recent data\cite{HFAG} which gives:
\bea
&& {\cal B}(B^{-}\to \rho^{-}\eta)= (5.4\pm 1.2)\times 10^{-6} , \nonumber\\  
&& {\cal B}(B^{-}\to \rho^{-}\eta')= (9.1^{+3.7}_{-2.8})\times 10^{-6} .
\label{etarhobr}
\eea
Since the above tree-dominated decays with $\eta, \eta'$ in the final state
are more sensitive to the $F^{B\to \eta} $ and $F^{B\to \eta'}$
form factor,  by increasing the $F^{B\to \eta'}$ form factor by
$40-50\%$ from the nonet symmetry value, one could bring 
${\cal B}(B^{-}\to \pi^{-}\eta')$, 
${\cal B}(B^{-}\to \rho^{-}\eta')$, and ${\cal B}(B^{-}\to K^{-}\eta')$, 
closer to the measured branching ratios. For example, by taking 
$F^{B\to \eta'}_{0}(0)=0.156$, one gets:
\bea
&& {\cal B}(B^{-}\to \pi^{-}\eta')= 3.888\times 10^{-6} , \nonumber\\  
&& {\cal B}(B^{-}\to K^{-}\eta')= 61.837\times 10^{-6} .
\label{bretaK}
\eea
which largely improves the prediction for ${\cal B}(B^{-}\to K^{-}\eta') $ 
but the predicted ${\cal B}(B^{-}\to \pi^{-}\eta')$ slightly
exceeds the HFAG new average, though consistent with the Babar value for
this mode. We note also the predicted  ${\cal B}(B^{-}\to \rho^{-}\eta')$
in \cite{Zhu4} approaches the measured value with the increased
form factor $F^{B\to \eta'}_{0}(0)=0.156 $ . As mentioned earlier, additional
source of enhancement of ${\cal B}(B^{-}\to K^{-}\eta') $ could come 
from a possible higher order $SU(3)$ breaking effects in the matrix
element $\langle 0|\bar{s}\,i\gamma_5 s|s\bar{s}\rangle $ for $\eta_{0} $.
Assuming a $15\%$ increase of this matrix element from its nonet value,
we would have  ${\cal B}(B^{-}\to K^{-}\eta')= 69.375\times 10^{-6}$, very
close to the measured value.

\section{CONCLUSION} 
We have shown that nonet symmetry for the pseudo-scalar meson mass term
implies  nonet symmetry for the pseudo-scalar density matrix element.
We then use  nonet symmetry for  
the pseudo-scalar density matrix element and  the $B\to \eta$, $B\to \eta'$
form factors to compute two-body charmless $B$ decays with
$\eta,\eta'$ in the final state. The discrepancy with experiment for 
tree-dominated decays with $\eta'$ in the final state indicates that the 
$F^{B\to \eta'}_{0}(0)$ form factors should be bigger than the nonet symmetry 
value by $40-50\%$. This  value together with a 
moderate annihilation contribution found in $\bar{B}^{0}\to K^{-}\pi^{+}$ 
decay, produces $B^{-}\to K^{-}\eta'$ branching ratio close to data. 
Our value for the $F^{B\to \eta'}_{0}(0)$ form factor supports the current 
calculations in PQCD and light-cone sum rules approach\cite{Li,Ball1}.
A possible increase by $15\%$ of the pseudo-scalar density matrix element
$\langle 0|\bar{s}\,i\gamma_5 s|s\bar{s}\rangle $ for $\eta_{0} $ would 
bring the predicted $B^{-}\to K^{-}\eta'$ branching ratio very close to
experiment. 

\bigskip

\begin{center}
{\bf Acknowledgments} \end{center}
I would like to thank R. Zwicky for reminding me of the current light-cone
sum rules values for the $B\to \pi$ and $B\to K$ form factors.
This work was supported in part by the EU 
contract No. MRTN-CT-2006-035482, "FLAVIAnet".


\end{document}